\documentstyle[prl,aps,floats,eqsecnum,preprint]{revtex}

\begin{document}

\title{Covariance properties and regularization of conserved currents in 
tetrad gravity}

\author{Yuri N.~Obukhov\footnote{yo@ift.unesp.br}}
\address {Instituto de F\'{\i}sica Te\'orica, Universidade Estadual Paulista,  
Rua Pamplona 145,  01405-900 S\~ao Paulo, Brazil}
\address{Department of Theoretical Physics, Moscow State University, 117234
Moscow, 
Russia}
\author{Guillermo F.~Rubilar\footnote{grubilar@udec.cl}}
\address {Instituto de F\'{\i}sica Te\'orica, Universidade Estadual Paulista,  
Rua Pamplona 145,  01405-900 S\~ao Paulo, Brazil}
\address{Departamento de F{\'{\i}}sica, Universidad de Concepci\'on,
Casilla 160-C, Concepci\'on, Chile}

\maketitle

\begin{abstract}
We discuss the properties of the gravitational energy-momentum 3-form within 
the tetrad formulation of general relativity theory. We derive the covariance 
properties of the quantities describing the energy-momentum content under 
Lorentz transformations of the tetrad. As an application, we consider the 
computation of the total energy (mass) of some exact solutions of Einstein's 
general relativity theory which describe compact sources with asymptotically 
flat spacetime geometry. As it is known, depending on the choice of tetrad 
frame, the formal total integral for such configurations may diverge. We 
propose a natural regularization method which yields finite values for the 
total energy-momentum of the system and demonstrate how it works on a number 
of explicit examples. 
\end{abstract}
\bigskip



\section{Introduction}

The problem of energy-momentum of the gravitational field belongs to 
the oldest in modern theoretical physics. The concepts of energy and 
momentum are fundamental ones in classical field theory. Within the 
general Lagrange-Noether approach, conserved currents arise from the 
invariance of the classical action under transformations of fields. In 
particular, energy and momentum are related to time and space translations. 
However, due to the geometric nature of the gravitational theory and 
because of the equivalence principle which identifies locally gravity 
and inertia, the definition of gravitational energy remains a problem.
In general, there are no symmetries in Riemannian manifolds that can
be used to generate the corresponding conserved energy-momentum currents.
It is possible, though, to associate energy and momentum to 
asymptotically flat gravitational field configurations. The history of
the problem and corresponding achievements is described in reviews 
\cite{Trautman,Faddeev,Szabados}, for example. 

There are several approaches to the study of the gravitational energy. 
The traditional one is to use the metric tensor as the field variable,
choose the local spacetime coordinates and construct the energy-momentum 
pseudotensor. Well known examples are the Einstein \cite{Ein}, 
Landau-Lifshitz \cite{LL}, Bergmann \cite{PB} expressions or their 
generalizations \cite{Goldberg}. A nice overview can be found in 
\cite{Nester1,Nester2}. Closely related is the Hamiltonian approach in 
which, after the choice of a (1+3)-decomposition of spacetime, one takes 
the spatial components of the metric together with the conjugate momenta 
as the basic variables \cite{adm,Faddeev,Blag}. The total energy of the 
asymptotically flat configurations is then defined by the surface (boundary) 
terms in the Hamiltonian \cite{RT}. A discussion of covariant Hamiltonian 
formulation of general relativity and of gauge gravity models is presented in 
\cite{Nester1,Nester2,CN99,Nester3,Hecht}. Of more recent developments 
it is worthwhile to mention the quasilocal approach in which the conserved 
quantities are associated to extended but finite domains of spacetime, 
see the extensive review and the literature in \cite{Szabados}. 

We will analyse the definition and the properties of the energy-momentum
in the framework of the tetrad formulation of general relativity. This 
approach was started with the work of M\o{}ller \cite{Moller}, for the 
early developments see also \cite{Pele,Kaempfer,Rodicev}. More recently 
the interest in the tetrad formulation of gravity has been revived 
\cite{Cho,Thirring,HSh79,Meyer,Muench97}. In particular, considerable 
efforts 
\cite{Maluf94,Maluf95,dAGP00,Itin,Maluf04,Maluf05b,Wallner,Dubois,Estabrook} 
have been devoted to the study of the properties of the energy-momentum 
densities of gravitating systems within the formalism of the so-called 
teleparallel equivalent of general relativity (TEGR). The latter can be 
formulated as a gauge theory of the spacetime translation group. The 
corresponding energy-momentum currents can be defined following the 
traditional gauge approach \cite{dAGP00,Itin}, or from a Hamiltonian 
formulation \cite{Maluf94}. These energy-momentum currents are naturally 
covariant under general coordinate transformations, invariant under gauge 
transformations, and transform covariantly under \textit{global} Lorentz 
transformations of the coframe (tetrad field) \cite{dAGP00}. On the other 
hand, they are {\it not covariant under local Lorentz transformations}. 
As a consequence, the corresponding energy and momentum densities and 
also the total conserved quantities in a given spacelike region are 
coordinate independent. However, they do depend on the chosen frame. It 
has been shown that the energy so defined agrees with the ADM energy 
\cite{adm,Faddeev} for asymptotically flat spacetimes \cite{Maluf95}. 
Furthermore, it also gives the correct Bondi energy \cite{Maluf04}. All 
the relevant computations have required the use of tetrads with appropriate 
asymptotic behavior at spatial infinity. On the other hand, there have 
been some attempts to study the properties and proper interpretation of 
the conserved quantities when computed in frames related by Lorentz 
transformations which are not global, see \cite{Maluf05b}. These studies 
are restricted to some particular solutions of the gravitational field 
equations and to quite specific choices of frames.

One of the aims of our paper is to systematically investigate the covariance
properties of the energy-momentum currents under changes of the tetrad. 
Accordingly, we derive the explicit transformation laws of the canonical 
current and of the related field momentum under finite general linear
and local Lorentz transformations. Although such a computation can be
done directly within the framework of the (purely) tetrad formulation, 
we find it more convenient to use the natural embedding of the tetrad 
gravity into the general scheme of the metric-affine gravity (MAG) models. 
The corresponding formalism was developed for the gauge approach to 
gravity on the basis of the general affine group \cite{HMMN95,Gronwald97} 
and it was applied recently to the analysis of the general teleparallel 
gravity models \cite{telemag}.

It turns out that the transformation of the field momentum and of the 
canonical energy-momentum are accompanied by a change of the Lagrangian.
The latter is shifted by a total derivative that effectively changes the
boundary conditions while leaving the field equations invariant. All the 
transformed energy-momentum currents are conserved like the original one, 
so in this way we derive a family of conserved energy-momenta parametrized 
by the elements of the local Lorentz group. This family is associated with 
a corresponding family of boundary conditions, in complete agreement 
with observations in \cite{Nester1}. 

{}From the physical point of view, a choice of a (co)frame can be interpreted 
as a choice of the reference system of an observer. It is obvious that the
observer's own dynamics, the state and structure of the corresponding 
reference 
system, can affect the physical measurements, including the determination of 
the energy and momentum of the gravitational systems. A particular example 
of the change of the total energy-momentum for a boosted observer is 
discussed in \cite{Maluf05b}. The knowledge of the covariance properties 
of the field momentum and the energy-momentum current is crucial for 
understanding the behavior of the total energy-momentum under change 
of reference frame. We demonstrate that for reference frames which are 
related by \textit{asymptotically global} Lorentz transformations, 
the total energy-momentum transforms as a 4-vector, as expected. 

The computation of the total energy in the framework of the tetrad formulation
has the obvious merit of general covariance. At all steps the choice of the 
spacetime coordinates is unimportant because of the use of exterior forms
which are coordinate invariant. However, an unfortunate choice of a (co)frame
may result in a formally infinite value of the total energy-momentum (this
may happen even for a flat spacetime). Accordingly, a regularization is
needed, in general. A particular regularization scheme was recently proposed
in \cite{Maluf05} which is based on the idea of subtracting a contribution of
a suitably introduced ``reference" spacetime geometry. Here we develop a 
different regularization recipe which is based on the {\it relocalization}
of the energy-momentum currents. At the same time, the regularized currents
become also covariant under the transformations of (co)frames. In this
sense the proposed scheme is similar to the covariant Hamiltonian approach
described in \cite{Nester1,Nester2,CN99,Nester3,Hecht}. 

Our general notations are as in \cite{HMMN95}. In particular, we use the 
Latin indices $i,j,\dots$ for local holonomic spacetime coordinates and the
Greek indices $\alpha,\beta,\dots$ label (co)frame components. Particular 
frame components are denoted by hats, $\hat 0, \hat 1$, etc. As usual,
the exterior product is denoted by $\wedge$, while the interior product of a 
vector $\xi$ and a $p$-form $\Psi$ is denoted by $\xi\rfloor\Psi$. The vector 
basis dual to the frame 1-forms $\vartheta^\alpha$ is denoted by $e_\alpha$ 
and they satisfy $e_\alpha\rfloor\vartheta^\beta=\delta^\beta_\alpha$. 
Using local coordinates $x^i$, we have $\vartheta^\alpha=h^\alpha_idx^i$ and 
$e_\alpha=h^i_\alpha\partial_i$. We define the volume 4-form by 
$\eta:=\vartheta^{\hat{0}}\wedge\vartheta^{\hat{1}}\wedge
\vartheta^{\hat{2}}\wedge\vartheta^{\hat{3}}$. Furthermore, with the help of 
the interior product we define $\eta_{\alpha}:=e_\alpha\rfloor\eta$, 
$\eta_{\alpha\beta}:=e_\beta\rfloor\eta_\alpha$, $\eta_{\alpha\beta\gamma}:=
e_\gamma\rfloor\eta_{\alpha\beta}$ which are bases for 3-, 2- and 1-forms 
respectively. Finally, $\eta_{\alpha\beta\mu\nu} = e_\nu\rfloor\eta_{\alpha
\beta\mu}$ is the Levi-Civita tensor density. The $\eta$-forms satisfy the
useful identities:
\begin{eqnarray}
\vartheta^\beta\wedge\eta_\alpha &=& \delta_\alpha^\beta\eta ,\\
\vartheta^\beta\wedge\eta_{\mu\nu} &=& \delta^\beta_\nu\eta_{\mu} -
\delta^\beta_\mu\eta_{\nu},\label{veta1}\\ \label{veta}
\vartheta^\beta\wedge\eta_{\alpha\mu\nu}&=&\delta^\beta_\alpha\eta_{\mu\nu
} + \delta^\beta_\mu\eta_{\nu\alpha} + \delta^\beta_\nu\eta_{\alpha\mu},\\
\label{veta2}
\vartheta^\beta\wedge\eta_{\alpha\gamma\mu\nu}&=&\delta^\beta_\nu\eta_{\alpha
\gamma\mu} - \delta^\beta_\mu\eta_{\alpha\gamma\nu} + \delta^\beta_\gamma
\eta_{\alpha\mu\nu} - \delta^\beta_\alpha\eta_{\gamma\mu\nu}.
 \end{eqnarray}
The line element $ds^2 = 
g_{\alpha\beta}\vartheta^\alpha\otimes\vartheta^\beta$ is defined by the 
spacetime metric $g_{\alpha\beta}$ of signature $(+,-,-,-)$. 

\section{Translational gauge theory: tetrad gravity}\label{sectetrad}

Einstein's general relativity (GR) theory admits a reformulation as a 
tetrad theory by replacing the metric with the (co)frame as the basic 
field variable. It is possible to interpret such a reformulation as a gauge 
theory of the spacetime translation group. In the usual Yang-Mills theory
of internal symmetry groups, the gauge field potential arises as a 1-form 
with values in the corresponding Lie algebra. In complete agreement 
with the general scheme, the translational gauge potential is a quartet of
1-forms $B^\alpha = B^\alpha_idx^i$, with Greek index labeling the four 
generators of the abelian group of the spacetime translations. In this 
sense, the theory is constructed along the same lines as electrodynamics 
is constructed as a gauge theory of the abelian one-dimensional group 
$U(1)$. Ultimately, such an approach introduces on the spacetime a coframe 
1-form $\vartheta^\alpha = dx^\alpha + B^\alpha$ that is invariant under 
local translations. {}From the gauge field potential $B^\alpha$ we 
define the \textit{field strength}
\begin{equation}
F^\alpha = dB^\alpha \equiv d\vartheta^\alpha. \label{F1}
\end{equation}
In geometrical terms, this is the anholonomity object of the coframe 
$\vartheta^\alpha$. 

Now, the Yang-Mills type Lagrangian 4-form for the tetrad $\vartheta^\alpha$ 
is constructed as follows as the sum of the quadratic invariants of the field
strength: 
\begin{equation}
\tilde{V}(\vartheta,d\vartheta) = -\,\frac{1}{2\kappa}\,F^\alpha\wedge{}^\star
\left({}^{(1)}F_{\alpha} - 2\,{}^{(2)}\!F_{\alpha} -\frac{1}{2}\,{}^{(3)}
\!F_{\alpha}\right). \label{V1}
\end{equation}
Here $\kappa=8\pi G/c^3$, and ${}^\star$ denotes the Hodge dual of the metric
$g_{\alpha\beta}$. The latter is assumed to be the flat Minkowski metric 
$g_{\alpha\beta} = o_{\alpha\beta} :={\rm diag}(+1,-1,-1,-1)$, and it is 
used also to raise and lower the Greek (local frame) indices. As it is 
well known, we can decompose the field strength $F^\alpha$ into the three 
irreducible pieces of the field strength:
\begin{eqnarray}\label{Fi1}
{}^{(1)}F^{\alpha}&:=&F^{\alpha}-{}^{(2)}F^{\alpha}-{}^{(3)}F^{\alpha},\\
{}^{(2)}F^{\alpha}&:=&\frac{1}{3}\,\vartheta^\alpha\wedge\left(e_\beta\rfloor
F^\beta\right),\label{Fi2}\\
{}^{(3)}F^{\alpha}&:=&\frac{1}{3}\,e^\alpha\rfloor\left(\vartheta^\beta\wedge
F_\beta\right),\label{Fi3}
\end{eqnarray}
i.e., the tensor part, the trace, and the axial trace, respectively. This 
Lagrangian is a particular case of the general 3-parameter Lagrangian 
quadratic 
in the field strength, see \cite{Gronwald97,HSh79,Itin,telemag} for details. 

In accordance with the general Lagrange-Noether scheme 
\cite{Gronwald97,HMMN95} we derive from (\ref{V1}) the translational gauge 
field momentum 2-form and the canonical energy-momentum 3-form:
\begin{eqnarray}
\tilde{H}_{\alpha} = -\,{\frac {\partial \tilde{V}} {\partial F^\alpha}} 
&=&\,{1\over \kappa}\,{}^\star\!\left({}^{(1)}F_{\alpha} - 2\,{}^{(2)}
F_{\alpha} - {1\over 2}\,{}^{(3)}F_{\alpha}\right), \label{defH} \\
\tilde{E}_\alpha = {\frac {\partial \tilde{V}} {\partial \vartheta^\alpha}}
&=& e_\alpha\rfloor \tilde{V} + (e_\alpha\rfloor F^\beta)\wedge 
\tilde{H}_\beta. \label{defE}
\end{eqnarray}
Accordingly, the variation of the total Lagrangian $L=\tilde{V} + 
L_{\rm mat}$ with respect to the tetrad results in the gravitational 
field equations
\begin{equation}
d\tilde{H}_\alpha-\tilde{E}_\alpha=\Sigma_\alpha, \label{fe1}
\end{equation}
with the canonical energy-momentum current 3-form of matter 
\begin{equation}
\Sigma_\alpha:=\frac{\delta L_{\rm mat}}{\delta \vartheta^\alpha}
\label{defsigma}
\end{equation}
as the source. 

Using (\ref{defH}) we can write the gravitational Lagrangian as
\begin{equation}
\tilde{V}=-\frac{1}{2}\,F^\alpha\wedge\tilde{H}_\alpha. \label{Vtilde}
\end{equation}

\subsection{Conserved quantities and relocalization}\label{secconsreloc}

The tetrad formulation allows for a straightforward definition of the 
conserved quantities. From (\ref{fe1}) we get
\begin{equation}
d\left(\tilde{E}_\alpha+\Sigma_\alpha\right)=0, \label{conserv0}
\end{equation}
We interpret (\ref{conserv0}) as the conservation law of the total (gravity
plus matter) energy-momentum. On the other hand, since $\tilde{E}_\alpha+
\Sigma_\alpha = d\tilde{H}_\alpha$, the 2-form of the field momentum
$\tilde{H}_\alpha$ plays the role of the corresponding superpotential. We 
then define the total 4-momentum of the system within a spacelike
hypersurface $S$ as the integral
\begin{equation}
\tilde{P}_\alpha=\int_S \left(\tilde{E}_\alpha+\Sigma_\alpha\right).
\label{defP}
\end{equation}
Or, using the field equations, 
\begin{equation}
\tilde{P}_\alpha=\int_{\partial S} \tilde{H}_\alpha.
\end{equation}
However, there is a certain freedom to define the local energy-momentum 
densities. The field equations allow for a \textit{relocalization} of the form
\begin{eqnarray}
\tilde{E}'_\alpha&=&\tilde{E}_\alpha+d\Psi_\alpha, \label{relocE}\\
\tilde{H}'_\alpha&=&\tilde{H}_\alpha+\Psi_\alpha. \label{relocH}
\end{eqnarray}
Here $\Psi_\alpha$ is an arbitrary covector-valued 2-form (24 independent 
components). Obviously, $d\tilde{H}'_\alpha-\tilde{E}'_\alpha\equiv
d\tilde{H}_\alpha-\tilde{E}_\alpha$.

Relocalizations can be related to boundary contributions to the gravitational
action. Suppose we change the Lagrangian by a total differential: $\tilde{V}'
=\tilde{V} + d\Psi$. The second term will only contribute with a boundary
integral to the action. We assume that the 3-form $\Psi$ depends only on 
the coframe and, possibly, on some external field $\Phi$, i.e. $\Psi=\Psi(
\vartheta,\Phi)$. Then we still have a first order Lagrangian $V'(\vartheta,
d\vartheta)$, the field equations remain unchanged, but we find a modified 
canonical field momentum and energy-momentum of the form (\ref{relocE}), 
(\ref{relocH}), with $\Psi_\alpha=-\frac{\partial \Psi}{\partial
\vartheta^\alpha}$. Thus, a boundary term induces a relocalization. 
It is worthwhile to note, though, that not every relocalization is derived
from a boundary term. An important example is the well-known relocalization
of Belinfante-Rosenfeld which yields a symmetric energy-momentum current.

\subsection{Local Lorentz transformations}

The tetrad formulation is explicitly generally covariant since the exterior
forms are invariant under the change of the spacetime coordinates. However,
none of the basic quantities (\ref{F1}), (\ref{defH}), (\ref{defE}) are 
covariant under the local Lorentz transformations of the coframe. Their 
transformation laws under a change of the coframe $\vartheta'^\alpha=
\Lambda^\alpha_{\ \beta}(x)\,\vartheta^{\beta}$ can be derived directly from 
the corresponding definitions. After a long calculation we find
\begin{eqnarray}
\tilde{V}(\vartheta') &=& \tilde{V}(\vartheta) - {\frac
1 {2\kappa}}\,d\left[(\Lambda^{-1})^\beta{}_\gamma
d\Lambda^\gamma{}_\alpha\wedge
\eta^\alpha{}_\beta\right], \label{Vprime}\\
\tilde{E}'_\alpha(\vartheta') &=&  (\Lambda^{-1})^\beta{}_\alpha
\tilde{E}_\beta(\vartheta) + d(\Lambda^{-1})^\beta{}_\alpha\wedge
\tilde{H}_\beta  \label{transE}\nonumber\\
&&\qquad -\,{\frac 1 {2\kappa}}\,d\left[(\Lambda^{-1})^\beta{}_\alpha
(\Lambda^{-1})^\nu{}_\gamma
d\Lambda^\gamma{}_\mu\wedge\eta_\beta{}^\mu{}_\nu\right]. \label{Eprime}\\
\tilde{H}'_\alpha(\vartheta') &=&  (\Lambda^{-1})^\beta{}_\alpha
\tilde{H}_\beta(\vartheta) - {\frac 1 {2\kappa}}
(\Lambda^{-1})^\beta{}_\alpha (\Lambda^{-1})^\nu{}_\gamma
d\Lambda^\gamma{}_\mu
\wedge\eta_\beta{}^\mu{}_\nu\,. \label{Hprime}
\end{eqnarray}
We will not give a `brute force' proof here, but instead we will provide a 
simple demonstration using the framework of metric-affine gravity in section 
\ref{flt}. To our knowledge, the explicit transformation law for the 
teleparallel conserved quantities was not known in the literature, and 
here it is computed for the first time.

Using (\ref{Eprime}) and (\ref{Hprime}) we can now easily verify that 
the left hand side of the field equations (\ref{fe1}) is covariant:  
$d\tilde{H}'_\alpha-\tilde{E}'_\alpha=(\Lambda^{-1})^\beta_{\ \alpha}
(d\tilde{H}_\beta-\tilde{E}_\beta)$. Taking into account the covariance
of the canonical energy-momentum (\ref{defsigma}) of matter, $\Sigma'_\alpha
=(\Lambda^{-1})^\beta{}_\alpha\Sigma_\beta$, we thus prove that the field 
equations (\ref{fe1}) are covariant under local Lorentz transformations.

For the infinitesimal case $\Lambda^{\beta}{}_\alpha=\delta_{\alpha}^{\beta}
+\varepsilon_{\ \alpha}^{\beta}$ with $\varepsilon^{\alpha\beta}=-
\varepsilon^{\beta\alpha}$, we have 
\begin{eqnarray}
\delta\tilde{V} &=& - {\frac 1 {2\kappa}}\,d\left[d\varepsilon^\beta
{}_\alpha\wedge\eta^\alpha{}_\beta\right]= {\frac 1 {2\kappa}}\,d\left[
d\varepsilon^{\alpha\beta}\wedge\eta_{\alpha\beta}\right], \\
\delta\tilde{E}_\alpha &=&  -\varepsilon^\beta{}_\alpha
\tilde{E}_\beta(\vartheta) -d\varepsilon^\beta{}_\alpha\wedge
\tilde{H}_\beta+\,\frac{1}{2\kappa}\,d\left[d\varepsilon^{\beta\gamma}
\wedge\eta_{\alpha\beta\gamma}\right],\\ 
\delta\tilde{H}_\alpha &=&  -\varepsilon^\beta{}_\alpha\tilde{H}_\beta(
\vartheta) +\,\frac{1}{2\kappa}\,d\varepsilon^{\beta\gamma}
\wedge\eta_{\alpha\beta\gamma}.
\end{eqnarray}

\subsection{Lorentz transformations of total conserved charges}

Since the field equation is covariant, both $\tilde{E}_\alpha 
+\Sigma_\alpha$ and $\tilde{E}'_\alpha +\Sigma'_\alpha$ are conserved. 
As a consequence, (\ref{Hprime}) implies that the total 
conserved quantity (\ref{defP}) will change to
\begin{equation}
\tilde{P}'_\alpha=\int_{\partial S}(\Lambda^{-1})^\beta{}_\alpha
\tilde{H}_\beta
-\frac{1}{2\kappa}\int_{\partial S}(\Lambda^{-1})^\beta{}_\alpha
(\Lambda^{-1})^\nu{}_\gamma\, 
d\Lambda^\gamma{}_\mu\wedge\eta_\beta{}^\mu{}_\nu. \label{P'}
\end{equation}
Let us assume that the Lorentz transformation $\Lambda^\beta{}_\alpha$ 
is {\it asymptotically global}, i.e., $d\Lambda^\beta{}_\alpha$ vanishes 
at the boundary $\partial S$, then
\begin{equation}
\tilde{P}'_\alpha=(\overline{\Lambda}^{-1})^\beta{}_\alpha
\int_{\partial S}\tilde{H}_\beta=(\overline{\Lambda}^{-1})^\beta{}_\alpha
\tilde{P}_\beta \,,
\end{equation}
where $\overline{\Lambda} = \Lambda\vline_{\partial S}$. Thus, the total 
energy-momentum transforms as a vector under these Lorentz transformations.

\subsection{Equivalence with general relativity}

The resulting theory is equivalent to Einstein's GR. In order to verify
this, we first need to recover the Riemannian connection and the 
corresponding Riemannian curvature. Although at the beginning we did not 
assume any connection structure on the spacetime manifold, it can be defined 
from the tetrad field. Indeed, given the coframe $\vartheta^\alpha$ 
together with its field strength $F^\alpha = d\vartheta^\alpha$, we 
construct the 1-form 
\begin{equation}
\tilde{\Gamma}_{\alpha\beta}:=\frac{1}{2}\left[e_\alpha\rfloor
F_\beta-e_\beta\rfloor F_\alpha-(e_\alpha\rfloor
e_\beta\rfloor F_\gamma)\wedge \vartheta^\gamma\right].\label{deftildeGamma}
\end{equation}
Direct inspection shows that this is a Lorentz connection. Under a local
Lorentz transformation of the coframe $\vartheta'^\alpha = \Lambda^\alpha
_{\ \beta}\vartheta^\beta$ it transforms as $\tilde{\Gamma}_\alpha^{\prime
\ \beta}=(\Lambda^{-1})^\mu_{\ \alpha}\Gamma_\mu^{\ \nu}\Lambda^\beta_{\ \nu} 
+ \Lambda^\beta_{\ \gamma}d(\Lambda^{-1})^\gamma_{\ \alpha}$. Moreover,
its torsion (covariant derivative of the coframe) is zero, 
\begin{equation}
\tilde{D}\vartheta^\alpha = d\vartheta^\alpha+\tilde{\Gamma}_\beta^{\
\alpha}\wedge\vartheta^\beta\equiv 0,\label{defGammariem}
\end{equation}
whereas the metric is covariantly constant $\tilde{D}g_{\alpha\beta} = 
- \tilde{\Gamma}_\alpha{}^\gamma g_{\gamma\beta} - \tilde{\Gamma}_\beta
{}^\gamma g_{\alpha\gamma} = 0$ in view of the skew symmetry of 
(\ref{deftildeGamma}). Consequently, this is indeed a Riemannian 
connection. 

We will denote all the Riemannian objects and operations by the tilde.
The Riemannian curvature 2-form is, as usual, $\tilde{R}_\alpha{}^\beta 
= d\tilde{\Gamma}_\alpha{}^\beta + \tilde{\Gamma}_\gamma{}^\beta\wedge
\tilde{\Gamma}_\alpha{}^\gamma$. 

The final proof of the equivalence of the tetrad field equations (\ref{fe1})
to the Einstein field equations relies on the geometric identity 
\begin{equation}
{}^\star\!\left({}^{(1)}F_{\alpha} - 2\,{}^{(2)}F_{\alpha} - {1\over 2}
\,{}^{(3)}F_{\alpha}\right) \equiv \frac{1}{2}\tilde{\Gamma}^{\beta\gamma}
\wedge\eta_{\alpha\beta\gamma}.
\end{equation}
One can directly verify this by making use of (\ref{deftildeGamma}) and of
the definitions of the irreducible parts (\ref{Fi1})-(\ref{Fi3}), see also
\cite{telemag,telemag2,PGrev}. Accordingly, we find for the gravitational
field momentum
\begin{equation}
\widetilde{H}_\alpha=\frac{1}{2\kappa}\tilde{\Gamma}^{\beta\gamma}
\wedge\eta_{\alpha\beta\gamma}.\label{Htilde}
\end{equation}
Using (\ref{defGammariem}) and (\ref{Htilde}) in (\ref{Vtilde}),  
the gravitational Lagrangian can be recast in the form
\begin{equation}
\tilde{V} = {\frac 1{2\kappa}}\,\tilde{\Gamma}_\lambda{}^\mu\wedge
\tilde{\Gamma}^{\lambda\nu}\wedge\eta_{\mu\nu}.\label{lagEH}
\end{equation}
Here we used the identity (\ref{veta}). Hence,
\begin{equation}
e_\alpha\rfloor\tilde{V} = {\frac 1 \kappa}\,\left(e_\alpha\rfloor
\tilde{\Gamma}_\lambda{}^\mu\right)\wedge\tilde{\Gamma}^{\lambda\nu}
\wedge\eta_{\mu\nu} + {\frac 1{2\kappa}}\,\tilde{\Gamma}_\lambda{}^\mu
\wedge\tilde{\Gamma}^{\lambda\nu}\wedge\eta_{\alpha\mu\nu}.
\end{equation}
On the other hand, using (\ref{defGammariem}) and (\ref{Htilde}), again with 
the help of (\ref{veta}), we find
\begin{equation}
\left(e_\alpha\rfloor F^\beta\right)\wedge\tilde{H}_\beta
= - {\frac 1 \kappa}\,\left(e_\alpha\rfloor\tilde{\Gamma}_\lambda
{}^\mu\right)\wedge\tilde{\Gamma}^{\lambda\nu}\wedge\eta_{\mu\nu} +
{\frac 1{2\kappa}}\,\tilde{\Gamma}_\alpha{}^\beta\wedge
\tilde{\Gamma}^{\mu\nu}\wedge\eta_{\beta\mu\nu}.
\end{equation}
As a result, we obtain the explicit form for the gravitational
energy-momentum (\ref{defE})
\begin{equation}
\tilde{E}_\alpha = {\frac 1 \kappa}\,\tilde{\Gamma}_{[\alpha}{}^\mu
\wedge\eta_{\beta]\mu\nu}\wedge\tilde{\Gamma}^{\beta\nu}.\label{Sparling}
\end{equation}
This is known as the Sparling 3-form \cite{Sparling1,Sparling2}.

Finally, let us compute the derivative of the gauge momentum (\ref{Htilde}).
Explicitly, we have
\begin{equation}
d\widetilde{H}_\alpha=\frac{1}{2\kappa}\left(d\tilde{\Gamma}^{\beta\gamma}
\wedge\eta_{\alpha\beta\gamma} - \tilde{\Gamma}^{\beta\gamma}
\wedge d\eta_{\alpha\beta\gamma}\right).\label{dHtilde}
\end{equation}
Since $d\eta_{\alpha\beta\gamma} = \eta_{\alpha\beta\gamma\nu}
d\vartheta^\nu = -\,\eta_{\alpha\beta\gamma\nu}\tilde{\Gamma}_\mu{}^\nu
\wedge\vartheta^\mu$, after using (\ref{veta2}), we find
\begin{equation}
d\widetilde{H}_\alpha=\frac{1}{2\kappa}\left(d\tilde{\Gamma}^{\beta\gamma}
+ \tilde{\Gamma}_\nu{}^\beta\wedge\tilde{\Gamma}^{\nu\gamma}\right)\wedge
\eta_{\alpha\beta\gamma} + {\frac 1 \kappa}\,\tilde{\Gamma}_{[\alpha}{}^\mu
\wedge\eta_{\beta]\mu\nu}\wedge\tilde{\Gamma}^{\beta\nu}.\label{dHtilde2}
\end{equation}
Using (\ref{Sparling}) and (\ref{dHtilde2}), we can rewrite 
the tetrad field equations (\ref{fe1}) as Einstein's equation
\begin{equation}
\frac{1}{2\kappa}\,\tilde{R}^{\beta\gamma}\wedge\eta_{\alpha\beta\gamma}
= \Sigma_\alpha.
\end{equation}

\section{Tetrad theory as a degenerate metric-affine gravity}

The tetrad theory can be naturally ``embedded" into the general framework
of the metric-affine gravity (MAG). In MAG, the gravitational field potentials 
are the metric, coframe and the linear connection  $g_{\alpha\beta}, 
\vartheta^\alpha, \Gamma_\alpha{}^\beta$. The corresponding gauge field 
strengths are the nonmetricity 1-form $Q_{\alpha\beta}$, the torsion 2-form 
\begin{equation}
T^\alpha = d\vartheta^\alpha+\Gamma_\beta^{\ \alpha}\wedge\vartheta^\beta,
\label{defT}
\end{equation}
and the curvature 2-form $R_\alpha{}^\beta$, respectively. Teleparallel
gravity arises as a special case of MAG when we impose the constraints of
vanishing curvature and nonmetricity:
\begin{eqnarray}
R_\alpha{}^\beta &=& d\Gamma_\alpha{}^\beta + \Gamma_\gamma
{}^\beta\wedge\Gamma_\alpha{}^\gamma = 0,\label{Rzero}\\
Q_{\alpha\beta} &=& -dg_{\alpha\beta} + \Gamma_\alpha{}^\gamma g_{\gamma\beta}
+ \Gamma_\beta{}^\gamma g_{\alpha\gamma} =0.\label{Qzero}
\end{eqnarray}
A Yang-Mills type Lagrangian can thus be constructed as a quadratic polynomial
of the torsion. The latter has three irreducible pieces, defined along
the same pattern as (\ref{Fi1})-(\ref{Fi3}). Accordingly, the general 
teleparallel model is based on a 3-parameter Lagrangian. This case was
considered in \cite{telemag}. Here, however, we confine our attention to the 
so-called teleparallel equivalent gravity model with the specific Lagrangian
\begin{equation} \label{V2}
V(g,\vartheta,\Gamma) = -\,{\frac 1 {2\kappa}}T^{\alpha}\wedge{}^\star\left(
{}^{(1)}T_{\alpha}- 2{}^{(2)}T_{\alpha} -{1\over 2}{}^{(3)}T_{\alpha}\right).
\end{equation}
This Lagrangian coincides with (\ref{V1}) when the connection is trivial, 
$\Gamma_\beta{}^\alpha=0$.

The MAG field equations are obtained from the variation of the total action
with respect to the metric, coframe and connection. The so-called 
0-th equation which comes from the metric variation is identically 
satisfied, whereas the so-called 1-st equation reads
\begin{equation}
DH_\alpha - E_\alpha = \Sigma_\alpha.\label{geq}
\end{equation}
Here $D$ denotes the covariant exterior derivative, i.e., $DH_\alpha=d
H_\alpha-\Gamma_\alpha^{\ \beta}\wedge H_\beta$. The translational momentum
and the canonical energy-momentum are, respectively: 
\begin{eqnarray}
H_{\alpha} = -\,{\frac {\partial V} {\partial T^\alpha}} &=&
\,{1\over \kappa}\,{}^\star\!\left({}^{(1)}T_{\alpha} - 2\,{}^{(2)}T_{\alpha}
- {1\over 2}\,{}^{(3)}T_{\alpha}\right),\label{Ha0}\\
E_\alpha = {\frac {\partial V} {\partial \vartheta^\alpha}} &=&
e_\alpha\rfloor V + (e_\alpha\rfloor T^\beta)\wedge H_\beta.\label{Ea0}
\end{eqnarray}

\subsection{Degeneracy of the MAG teleparallel equivalent model}

The specific feature of the MAG model (\ref{V2}) is that it is {\it 
degenerate} 
in the following sense. Let us fix the metric and the coframe and make a
shift of the connection. More exactly, let us consider the transformation 
of the gravitational variables $(g_{\alpha\beta},\vartheta^\alpha,
\Gamma_\beta^{\ \alpha})\rightarrow(g'_{\alpha\beta},\vartheta'^\alpha,
\Gamma_\beta^{\prime\ \alpha})$ of the form
\begin{equation}\label{shift}
g'_{\alpha\beta} = g_{\alpha\beta},\qquad\vartheta'^\alpha=\vartheta^\alpha,
\qquad \Gamma'_\beta{}^\alpha=\Gamma_\beta{}^\alpha + \Psi_\beta{}^\alpha.
\end{equation}
We assume that the 1-form $\Psi_\beta{}^\alpha$ is such that 
the teleparallel constraints (\ref{Rzero}) and (\ref{Qzero}) are preserved. 
Then from (\ref{Qzero}) we immediately see that $\Psi_\beta{}^\alpha$ 
is Lorentz-valued, i.e. $\Psi^{\alpha\beta}=-\Psi^{\beta\alpha}$. 

Under the shift (\ref{shift}) with any $\Psi_\beta{}^\alpha$ that preserves 
the 
teleparallel conditions, the Lagrangian (\ref{V2}) changes by a total 
derivative: 
\begin{equation}
V(g,\vartheta,\Gamma') = V(g,\vartheta,\Gamma) - \frac{1}{2\kappa}\,d\left(
\Psi^{\alpha\beta}\wedge\eta_{\alpha\beta}\right). \label{magic}
\end{equation}

The proof of this fact relies on the following geometric identity 
between the contractions of the irreducible pieces of the torsion and of the
nonmetricity \cite{effective,HMMN95}:
\begin{eqnarray}
&& R_{\alpha\beta}\wedge\eta^{\alpha\beta} + {}^{(1)}T^{\alpha}\wedge
{}^{\star(1)}T_{\alpha} - 2{}^{(2)}T^{\alpha}\wedge{}^{\star(2)}T_{\alpha} -
{1\over 2}{}^{(3)}T^{\alpha}\wedge{}^{\star(3)}T_{\alpha}\nonumber\\
&& - {}^{(2)}Q_{\alpha\beta}\wedge\vartheta^{\beta}\wedge
{}^{\star(1)}T^{\alpha} + 2{}^{(3)}Q_{\alpha\beta}\wedge
\vartheta^{\beta}\wedge{}^{\star(2)}T^{\alpha} + 2{}^{(4)}Q_{\alpha\beta}
\wedge\vartheta^{\beta}\wedge{}^{\star(2)}T^{\alpha}\nonumber\\
&&- {1\over 4}{}^{(1)}Q_{\alpha\beta}\wedge{}^{\star(1)}Q^{\alpha\beta}
+{1\over 2}{}^{(2)}Q_{\alpha\beta}\wedge{}^{\star(2)}Q^{\alpha\beta}
+{1\over 8}{}^{(3)}Q_{\alpha\beta}\wedge{}^{\star(3)}Q^{\alpha\beta}
\nonumber\\
&&-{3\over 8}{}^{(4)}Q_{\alpha\beta}\wedge{}^{\star(4)}Q^{\alpha\beta}
- ({}^{(3)}Q_{\alpha\gamma}\wedge\vartheta^{\alpha})\wedge
{}^\star({}^{(4)}Q^{\beta\gamma}\wedge\vartheta_{\beta})\nonumber\\
&& \equiv \widetilde{R}_{\alpha\beta}\wedge\eta^{\alpha\beta} - 
d\left[\vartheta^{\alpha}\wedge{}^\star\left(2T_{\alpha} - Q_{\alpha\beta}
\wedge\vartheta^{\beta}\right)\right].\label{identity}
\end{eqnarray}
With the help of the constraints (\ref{Rzero}) and (\ref{Qzero}), this 
identity recasts the Lagrangian (\ref{V2}) into
\begin{equation}
V(g,\vartheta,\Gamma) = {\frac 1 {2\kappa}}\,\widetilde{R}_{\alpha\beta}
\wedge\eta^{\alpha\beta} + {\frac 1 \kappa}\,d\left[\vartheta_{\alpha}
\wedge{}^\star\left(d\vartheta^\alpha + \Gamma_\beta{}^\alpha\wedge
\vartheta^\beta\right)\right].\label{Rscal}
\end{equation}
Subtracting from this the same relation written for a shifted connection 
$\Gamma'=\Gamma+\Psi$, we prove (\ref{magic}) after some straightforward
algebra.

The property (\ref{magic}) is only valid for the particular combination of 
the squares of the irreducible pieces of torsion in (\ref{V2}), and it is
violated for the general 3-parameter teleparallel Lagrangians. The special
feature (\ref{magic}) has an important consequence: the connection is 
completely undetermined in the MAG form of the teleparallel equivalent 
gravity. Strictly speaking, this makes the particular model (\ref{V2}) 
a non-viable theory within the MAG framework, since the field equations 
do not determine all of the dynamical variables ($g, \vartheta$ and 
$\Gamma$). As a result, matter with spin (in general, with hypermomentum) 
cannot be consistently coupled to the gravitational field in this framework 
\cite{telemag,telemag2,Leclerc1}. This degeneracy of the MAG scheme should
be compared to the tetrad formulation of Sec.~\ref{sectetrad} that was 
shown to be equivalent to GR in a dynamically consistent way.

\subsection{Relation between MAG and tetrad objects}

Despite the degeneracy mentioned, it is very useful to consider the tetrad 
theory embedded into MAG because, in such an extended framework, the 
gravitational field Lagrangian 4-form $V$ is, by construction, invariant 
under the local linear transformations ($g_{\alpha\beta},\vartheta^\alpha,
\Gamma_\beta{}^\alpha)\rightarrow (g'_{\alpha\beta},\vartheta'^\alpha,
\Gamma'_\beta{}^\alpha)$ given by
\begin{eqnarray}
g'_{\alpha\beta} &=& (L^{-1})^\mu_{\ \alpha}(L^{-1})^\nu_{\ \beta}
\,g_{\mu\nu},\label{mettrans}\\ 
\vartheta'^\alpha &=& L^\alpha_{\ \beta}\vartheta^\beta,\qquad
\Gamma_\alpha^{\prime\ \beta} = (L^{-1})^\mu_{\ \alpha}\Gamma_\mu^{\
\nu}L^\beta_{\ \nu} + L^\beta_{\ \gamma}
d(L^{-1})^\gamma_{\ \alpha},\label{cofcontrans}
\end{eqnarray}
$ L^\alpha_{\ \beta}(x)\in GL(4,R)$. The constraints (\ref{Rzero})
and (\ref{Qzero}) are preserved by these transformations.

We will use this fact for deriving the covariance properties of the basic 
objects in the tetrad gravity. In particular, we are interested in the
properties of the field momentum and the energy-momentum (\ref{defH}) and 
(\ref{defE}) under local linear and Lorentz transformations. Recall that
field strength $F^\alpha = d\vartheta^\alpha$ is not a covariant object 
under these transformations. As a result, the transformation of 
$\tilde{H}_\alpha$ and $\tilde{E}_\alpha$ is nontrivial. But on the other 
hand, the MAG counterparts of the field momentum and the energy-momentum 
derived from the Lagrangian (\ref{V2}), namely, (\ref{Ha0}) and 
(\ref{Ea0}) are explicitly {\it covariant}. 

This offers a simple way to compute the transformation laws of tetrad
quantities of Sec.~\ref{sectetrad}, using the known covariant 
transformation properties of the fields in the MAG picture. For this we
need to establish the explicit relations between the objects in MAG and
tetrad framework. 

To begin with, let us use the local linear transformations and bring the 
metric to the constant Minkowski values, $g_{\alpha\beta}=o_{\alpha\beta}$.
The remaining linear transformations which leave the metric invariant, are
just the local Lorentz transformations. In other words, the theory falls
naturally into the geometrical framework of the Poincar\'e gauge theory. 
Now we recall that the metric-compatible 
connection $\Gamma_\alpha{}^\beta$ can be decomposed into the Riemannian 
and post-Riemannian parts as
\begin{equation}
\Gamma_\alpha{}^\beta = \tilde{\Gamma}_\alpha{}^\beta -
K_\alpha{}^\beta .\label{gagaK}
\end{equation}
Here $\tilde{\Gamma}_\alpha{}^\beta$ is the purely Riemannian connection
(\ref{deftildeGamma}) and $K_\alpha{}^\beta$ is the contortion which is
related to the torsion via the identity
\begin{equation}
T^\alpha=K^\alpha{}_\beta\wedge\vartheta^\beta.\label{contor}
\end{equation}
Then one can show that due to geometric identities \cite{PGrev}, the gauge
momentum (\ref{Ha0}) can be written as
\begin{equation}
H_\alpha ={\frac 1 {2\kappa}}\,K^{\mu\nu}\wedge\eta_{\alpha\mu\nu}.\label{H0K}
\end{equation}
As a result, the Lagrangian (\ref{V2}) is recast as
\begin{equation}
V = -\,{\frac 12}\,T^\alpha\wedge H_\alpha = -\,{\frac 1{4\kappa}}\,T^\alpha
\wedge K^{\mu\nu}\wedge\eta_{\alpha\mu\nu}.\label{Vlag}
\end{equation}
Using (\ref{gagaK}) and (\ref{H0K}) we find the relation between the field
momenta in MAG and tetrad pictures:
\begin{equation}
H_\alpha = \tilde{H}_\alpha - {\frac 1 {2\kappa}}\,\Gamma^{\mu\nu}
\wedge\eta_{\alpha\mu\nu}.\label{HH}
\end{equation}
The direct computation for the energy-momentum 3-form yields 
\begin{equation}
E_\alpha = \tilde{E}_\alpha - \Gamma_\alpha{}^\beta\wedge H_\beta
- {\frac 1{2\kappa}}\,d\left(\Gamma^{\mu\nu}\wedge
\eta_{\alpha\mu\nu}\right).\label{EE}
\end{equation}
As a result, we verify that the left-hand side of the field equation
(\ref{geq}) also depends on the Riemannian variables only:
\begin{equation}
DH_\alpha - E_\alpha = dH_\alpha - \Gamma_\alpha{}^\beta\wedge H_\beta
- E_\alpha = d\tilde{H}_\alpha - \tilde{E}_\alpha.\label{geqR}
\end{equation}

Finally, directly form the identity (\ref{magic}), (\ref{Rscal}) we read 
off the relation between the Lagrangians in the MAG and tetrad pictures:
\begin{equation}
V = \tilde{V} - {\frac 1{2\kappa}}\,d\left(\Gamma^{\mu\nu}\wedge
\eta_{\mu\nu}\right).\label{VV}
\end{equation}

\subsection{Finite Lorentz transformations for the tetrad theory}\label{flt}

We will now find the behavior of the quantities defined in the tetrad
formulation under local Lorentz transformations. The MAG formulation 
is, by construction, covariant under the general linear transformations 
(\ref{mettrans})--(\ref{cofcontrans}). In the particular case when the 
metric components are fixed to be $g_{\alpha\beta}=o_{\alpha\beta}$, 
the symmetry is reduced to local Lorentz transformations of the frame 
$e'_\alpha = (\Lambda^{-1})^\beta{}_\alpha\, e_\beta$, and of the
coframe and connection:
\begin{eqnarray}
\vartheta'{}^\alpha &=& \Lambda^\alpha{}_\beta\,\vartheta^\beta,\label{vv}\\
\Gamma'{}_\alpha{}^\beta &=& (\Lambda^{-1})^\mu{}_\alpha\,\Gamma_\mu{}^\nu
\Lambda^\beta{}_\nu + \Lambda^\beta{}_\gamma\, 
d(\Lambda^{-1})^\gamma{}_\alpha.\label{gg}
\end{eqnarray}
In particular, the Lagrangian (\ref{V2}) is invariant,
\begin{equation}
V(\vartheta',\Gamma') = V(\vartheta,\Gamma),\label{Vinv}
\end{equation}
and the gauge momentum (\ref{Ea0})  as well as  the energy-momentum 3-form 
(\ref{Ea0}) are covariant,
\begin{eqnarray}
H'{}_\alpha(\vartheta',\Gamma') &=& (\Lambda^{-1})^\beta{}_\alpha\,
H_\beta(\vartheta,\Gamma),\label{Hcov} \\
E'{}_\alpha(\vartheta',\Gamma') &=& (\Lambda^{-1})^\beta{}_\alpha\,
E_\beta(\vartheta,\Gamma),\label{Ecov}
\end{eqnarray}
under the transformations (\ref{gg}).

We are, however, interested in the transformation properties of the tetrad 
formulation. Then we notice that the tetrad configuration
\begin{equation}
\left\{g_{\alpha\beta}=o_{\alpha\beta}, \quad \vartheta'{}^\alpha = 
\Lambda^\alpha{}_\beta\vartheta^\beta,
\quad \Gamma'{}_\alpha{}^\beta = 0\right\}\label{configR}
\end{equation}
is related by the Lorentz transformation (\ref{gg}) to the
configuration
\begin{equation}\label{configRC}
\left\{g_{\alpha\beta}=o_{\alpha\beta},\quad \vartheta^\alpha, \quad 
\Gamma_\alpha{}^\beta =
(\Lambda^{-1})^\beta{}_\gamma d\Lambda^\gamma{}_\alpha\right\}.
\end{equation}
Consequently, using (\ref{VV}) and (\ref{EE}), we have for the tetrad objects
\begin{equation}
\tilde{V}(\vartheta') = V(\vartheta',\Gamma'),\quad
\tilde{E}'_\alpha(\vartheta') = E'_\alpha(\vartheta',\Gamma').
\end{equation}
Finally, combining this with the equations (\ref{Vinv}), (\ref{Hcov}), 
(\ref{Ecov}), and again using (\ref{VV}), (\ref{HH}) and (\ref{EE}),  
we find the explicit transformation of the tetrad formulation, i.e., 
expressions (\ref{Vprime})-(\ref{Hprime}).

\subsection{Conformal symmetry and the trace of energy-momentum}

It has been noticed \cite{dAGP00,Itin,Maluf05b} that the energy-momentum 
current $\tilde{E}_\alpha$ is traceless. Here we show that this property 
can be understood as a consequence of the conformal symmetry of the 
corresponding MAG theory. Normally, the vanishing of the trace of the 
energy-momentum current is related to the conformal symmetry of the model, 
recall the Maxwell electrodynamics and the Klein-Gordon theory of a scalar 
field with a conformal coupling. The same holds true for 
teleparallel gravity. Namely, the vanishing of the trace of the 
energy-momentum, i.e. $\vartheta^\alpha\wedge E_\alpha=0$, is a direct
consequence of the invariance of the MAG Lagrangian (\ref{V2}) under 
conformal transformations. The latter are recovered as the 
one-parameter abelian subgroup of the general linear transformations
(\ref{mettrans}), (\ref{cofcontrans}) defined by the diagonal matrices
of the form $L^\alpha{}_\beta = e^{\lambda(x)}\,\delta^\alpha_\beta$.

Consider the infinitesimal conformal transformations of the MAG fields:
\begin{equation}
\delta g_{\alpha\beta} = -2\lambda\,g_{\alpha\beta}, \qquad 
\delta \vartheta^\alpha = \lambda\, \vartheta^\alpha ,\qquad
\delta \Gamma_\alpha^{\ \beta} = -d\lambda\,\delta^\beta_\alpha.
\label{conftrans}
\end{equation}
As a consequence,
\begin{equation}
\delta T^\alpha=\lambda\, T^\alpha.
\end{equation} 
The corresponding  variation of the teleparallel Lagrangian $V=V(g,
\vartheta,T)$ then can be written as
\begin{eqnarray}
\delta V&=&\delta g_{\alpha\beta}\frac{\partial V}{\partial 
g_{\alpha\beta}}+\delta \vartheta^\alpha\wedge\frac{\partial V}
{\partial\vartheta^\alpha}+\delta T^\alpha\wedge\frac{\partial V}{\partial 
T^\alpha}\\
&=&-2\lambda g_{\alpha\beta}\frac{\partial V}{\partial 
g_{\alpha\beta}}+\lambda \vartheta^\alpha\wedge\frac{\partial V}
{\partial\vartheta^\alpha}+\lambda T^\alpha\wedge\frac{\partial V}{\partial 
T^\alpha} \\
&=&\lambda\left[-2g_{\alpha\beta}\frac{\partial V}{\partial g_{\alpha\beta}}+ 
\vartheta^\alpha\wedge E_\alpha-T^\alpha\wedge H_\alpha\right] .
\label{deltaV}
\end{eqnarray}
In order to find the derivative of the Lagrangian with respect to the metric, 
we rewrite (\ref{Vlag}) as $V=-\frac{1}{2}g_{\mu\nu}T^\mu\wedge H^\nu$ and 
notice that besides the explicit first factor, the metric also enters the 
Hodge duality operator used in $H^\alpha={}^\star \phi^\alpha$, with 
$\phi^\alpha={1\over \kappa}\,\left({}^{(1)}T^{\alpha} - 2\,{}^{(2)}T^{\alpha}
- {1\over 2}\,{}^{(3)}T^{\alpha}\right)$, see (\ref{Ha0}). As a result, the 
derivative reads: 
\begin{equation}
\frac{\partial V}{\partial g_{\alpha\beta}}=-\frac{1}{2}T^{(\alpha}\wedge 
H^{\beta)}-\frac{1}{2}g_{\mu\nu}T^\mu\wedge \frac{\partial H^\nu}{\partial 
g_{\alpha\beta}}. \label{parV}
\end{equation} 
We can calculate the last term with the help of the ``master formula" 
derived in \cite{MGH98} for the variation of Hodge duals of arbitrary 
forms. From  Eq. (33) of \cite{MGH98} we have the identity
\begin{equation}
\frac{\partial ({}^\star\phi)}{\partial 
g_{\alpha\beta}}=\vartheta^{(\alpha}\wedge ( e^{\beta)} \rfloor{}^\star 
\phi ) -\frac{1}{2}\,g^{\alpha\beta}\,{}^\star \phi ,
\end{equation} 
for any metric-independent $p$-form $\phi$. Hence
\begin{equation}
\frac{\partial H^\nu}{\partial g_{\alpha\beta}}=\vartheta^{(\alpha}\wedge 
( e^{\beta)} \rfloor H^\nu) -\frac{1}{2}\,g^{\alpha\beta}\,H^\nu .
\label{parstar}
\end{equation} 
Substituting (\ref{parstar}) in (\ref{parV}) and then in (\ref{deltaV}), we 
obtain
\begin{eqnarray}
\delta V&=&\lambda\left[T^\alpha\wedge H_\alpha+ \vartheta^\alpha\wedge 
E_\alpha-T^\alpha\wedge H_\alpha\right]\\
&=&\lambda\,\vartheta^\alpha\wedge E_\alpha.
\end{eqnarray} 
Thus, the invariance $\delta V =0$ of the Lagrangian under the
transformations (\ref{conftrans}) implies the vanishing of the trace 
of the canonical energy-momentum current $E_\alpha$. Moreover, using 
(\ref{EE}), (\ref{defT}), (\ref{Vtilde}), (\ref{V2}), (\ref{VV}) and 
some straightforward algebra, one can show that $\vartheta^\alpha\wedge
E_\alpha=\vartheta^\alpha\wedge \tilde{E}_\alpha$, so that also the 
trace of $\tilde{E}_\alpha$ vanishes. Similar arguments were recently 
presented in \cite{Itin}.

\section{Regularization via the relocalization}

Now we return to the tetrad formulation, in which the coframe 
$\vartheta^\alpha$ (the translational gauge potential) is the only field
variable. The corresponding gauge field strength is the anholonomity 2-form
$F^\alpha =d\vartheta^\alpha$. In general, as usual in classical field theory, 
one expects a compact-object configuration to have trivial asymptotic
values of the field strength at spatial infinity, and the integral conserved 
quantities to have finite values.

In the tetrad formulation, however, one can encounter situations when the 
field strength $F^\alpha$ and the Riemannian connection $\tilde{\Gamma}_\beta
{}^\alpha$ constructed from it, see (\ref{deftildeGamma}), do not vanish at 
spatial infinity. This can lead to infinite values of the energy-momentum 
integral even for the flat spacetime geometry. Let us consider this case in 
more detail, taking as an example the flat Minkowski spacetime in the 
spherical coordinate system $(t,r,\theta,\varphi)$. Let us choose the coframe 
$\vartheta^\alpha$ as
\begin{equation}
\vartheta^{\hat{0}} = cdt,\qquad \vartheta^{\hat{1}} = dr,
\qquad\vartheta^{\hat{2}}
= rd\theta,\qquad\vartheta^{\hat{3}} = r\sin\theta d\varphi.\label{cofM1}
\end{equation}
The field strength $F^\alpha = d\vartheta^\alpha$ is nontrivial everywhere, 
also at the spatial infinity ($t=$const, $r\rightarrow\infty$). The 
nonvanishing components of the corresponding Riemannian connection read
\begin{equation}\label{flatGam}
\tilde{\Gamma}_{\hat{1}}{}^{\hat{2}}
= d\theta,\qquad \tilde{\Gamma}_{\hat{1}}{}^{\hat{3}} = \sin\theta d\varphi,
\qquad \tilde{\Gamma}_{\hat{2}}{}^{\hat{3}} = \cos\theta d\varphi.
\end{equation}
Consider the 2-dimensional surface $\partial S = \left\{r = R,\theta,
\varphi\right\}$ as a spatial boundary. Then we find for the total energy 
at a fixed time 
\begin{equation}
\tilde{P}_{\hat{0}} = \int_{\partial S}\tilde{H}_{\hat{0}} = 
-\,{\frac {2R}{\kappa}}\int_{\partial S}\sin\theta\,
d\theta\wedge d\varphi = -\,{\frac {8\pi R}\kappa},
\end{equation}
which diverges in the limit of $R\rightarrow\infty$.

However, let us recall that the energy-momentum current and the field
momentum (which plays the role of a superpotential) are defined up to 
relocalizations. In the present case, $\tilde{H}_\alpha$ diverges at spatial 
infinity while the components of the Riemannian connection are finite in 
this limit. We then ``regularize'' the total energy by performing a 
relocalization with $\Psi_\alpha=-\frac{1}{2\kappa}\overline{\Gamma
}^{\mu\nu}\wedge\eta_{\alpha\mu\nu}$, as described in 
Sec.~\ref{secconsreloc}. The relocalized momentum then is finite at 
spatial infinity and the total energy-momentum vanishes, as expected.

Now we are in a position to formulate a general recipe to find finite
conserved quantities for arbitrary orthonormal frames describing nontrivial
spacetime geometries. As a first step, we check whether the coframe 
$\vartheta^\alpha$ is asymptotically holonomic or not. In the latter case, 
the field strength $F^\alpha$ has a nontrivial limit $\overline{F}^\alpha 
:= \left. F^\alpha\right|_{\partial S}$. Next, we find the limit of the 
Riemannian connection of the frame $\vartheta^\alpha$ at spatial infinity,
\begin{equation}
\overline{{\Gamma}}{}_\alpha^{\ \beta}:=\left.\tilde{\Gamma}_\alpha^{\ 
\beta}\right|_{\partial S}. \label{fc}
\end{equation}
In case this limit is nontrivial, we use this connection (defined in the
interior domain by a continuous extension from the boundary) to perform 
a relocalization along the lines of Sec.~\ref{secconsreloc} with the boundary 
term defined by the 3-form $\Psi=\frac{1}{2\kappa}\overline{\Gamma}^{\mu\nu}
\wedge\eta_{\mu\nu}$. The latter depends on the original tetrad {\it and} on 
the connection $\overline{{\Gamma}}{}_\alpha^{\ \beta}$ (which plays the role
of the external field $\Phi$). Then $\frac{\partial\Psi}{\partial\
vartheta^\alpha} =\frac{1}{2\kappa}\overline{\Gamma}^{\mu\nu}\wedge
\eta_{\alpha\mu\nu}$ and the relocalized momentum reads
\begin{equation}
\widehat{H}_\alpha=\tilde{H}_\alpha-\frac{1}{2\kappa}\,\overline{{\Gamma}}
{}^{\mu\nu}\wedge\eta_{\alpha\mu\nu}=\frac{1}
{2\kappa}\Delta\Gamma^{\mu\nu}\wedge\eta_{\alpha\mu\nu} ,\label{Hreg}
\end{equation}
were $\Delta\Gamma^{\mu\nu}:=\tilde{\Gamma}^{\mu\nu}-\overline{\Gamma}^{\mu
\nu}$. This new momentum is similar to the ``improved'' covariant 
superpotentials considered in \cite{CN99}. The new conserved quantities,
\begin{equation}
\widehat{P}_\alpha = \int_{\partial\Sigma}\widehat{H}_\alpha, \label{Pa}
\end{equation}
are similar to the ``regularized" energy-momentum expressions proposed in 
\cite{Maluf05}. There are, however, important differences between the two 
regularization schemes. The regularization in \cite{Maluf05} is defined 
as a subtraction of a term constructed from a tetrad that describes a 
``background spacetime", different from the spacetime geometry of the 
original system. In contrast, in our formalism the original tetrad remains
the only frame, no ``reference frame" is introduced, but instead a 
Lorentz connection shows up as an additional structure. 

In a sense, the tetrad theory is effectively extended to the MAG framework. 
In particular, one can straightforwardly verify that the regularized 
Lagrangian is \textit{invariant} under local Lorentz transformations, 
whereas the regularized momentum is covariant.

The connection $\overline{\Gamma}^{\mu\nu}$ which we use for regularization, 
is not unique. Its role is to ``kill" the nontrivial (sometimes divergent)
behavior of the field momentum $\tilde{H}_\alpha$ at spatial infinity, 
providing a better spatial asymptotics of the regularized 
$\widehat{H}_\alpha$. In other words, the crucial condition is that 
$\Delta\Gamma^{\mu\nu}$ must vanish at spatial infinity. This still 
leaves some freedom to choose $\overline{\Gamma}^{\mu\nu}$, as we will 
illustrate with examples below. Different choices or $\overline{\Gamma
}^{\mu\nu}$ will lead to different total conserved quantities, in general. 

As a final remark, it is worthwhile to note that one can further use the 
resulting effective MAG scheme as follows. We consider flat Lorentz 
connections $\overline{\Gamma}$ (teleparallel structure), hence a local 
Lorentz matrix $\Lambda$ exists such that $\overline{\Gamma} = \Lambda^{-1}
d\Lambda$. We then can define a new transformed coframe $\vartheta' = 
\Lambda\vartheta$, that {\it by construction} does not require 
``regularization". Such a regular tetrad is obviously not unique, 
since $\Lambda$ is determined up to a global Lorentz matrix factor.

\subsection{Example 1: Kerr-Newman solution}\label{examples}

Let us now apply our method to some explicit computations of the total
energy-momentum of asymptotically flat configurations. In the first 
example, we consider the Kerr-Newman solution. We choose the spherical 
(Boyer-Lindquist) local coordinate system $(t,r,\theta,\varphi)$, and
write the coframe as \cite{HO03,PSH95}
\begin{eqnarray}
\vartheta^{\hat 0} &=& \sqrt{\frac{\Delta}{\Sigma}}\left[ c\,dt-a\sin^2\theta
\,d\varphi\right],\label{KNcof0} \\
\vartheta^{\hat 1} &=& \sqrt{\frac{\Sigma}{\Delta}}\, dr,\label{KNcof1} \\
\vartheta^{\hat 2} &=& \sqrt{\Sigma}\, d\theta,\label{KNcof2} \\
\vartheta^{\hat 3} &=& \frac{\sin\theta}{\sqrt{\Sigma}}\left[ -ac\,dt
+(r^2+a^2)\,d\varphi\right].\label{KNcof3}
\end{eqnarray}
Here the functions and constants are defined by
\begin{eqnarray}
\Delta&:=& r^2 + a^2 - 2mr + q^2,\\
\Sigma&:=& r^2 + a^2\cos^2\theta,\\
m &:=& \frac{GM}{c^2},\qquad q^2 := \frac{GQ^2}
{4\pi\varepsilon_0c^4}.\label{mq}
\end{eqnarray}
In accordance with our scheme, we now have to compute the field strength 
$F^\alpha$, the Riemannian connection $\tilde{\Gamma}_\alpha^{\ \beta}$ and 
the momentum $\tilde{H}_\alpha$. Since we need these quantities only
 asymptotically, we will express our results as a power series of
 $\frac{1}{r}$. We obtain the asymptotic behavior of the coframe components:
\begin{eqnarray}
\vartheta^{\hat 0} &=& \left(1 - {\frac m r}\right)\,cdt - a\sin^2\theta
\left(1 - {\frac m r}\right)\,d\varphi + \cdots, \\
\vartheta^{\hat 1} &=& \left(1 + {\frac m r}\right)\,dr + \cdots,\\
\vartheta^{\hat 2} &=& \left(r + {\frac {a^2\cos^2\theta}{2r}}\right)d\theta
+ \cdots, \\
\vartheta^{\hat 3} &=& -\,{\frac {a\sin\theta} r}\,cdt + \left[r + {\frac
{a^2}{2r}}\left(1 + \sin^2\theta\right)\right]\sin\theta\,d\varphi + \cdots.
\end{eqnarray}
The asymptotic behavior of their derivatives $F^\alpha =
d\vartheta^\alpha$ reads
\begin{eqnarray}
F^{\hat 0} &=& -\,{\frac m {r^2}}\,cdt\wedge dr - {\frac
{a^2\sin\theta\cos\theta}
{r^2}}\,cdt\wedge d\theta - {\frac {ma\sin^2\theta}{r^2}}\,dr\wedge d\varphi
\nonumber\\ && -\,2a\left(1 - {\frac m r}\right)\sin\theta\cos\theta
\,d\theta\wedge d\varphi + \cdots ,\label{F0KN}\\
F^{\hat 1} &=& {\frac {a^2\sin\theta\cos\theta}{r^2}}\,dr\wedge d\theta +
\cdots,\\
F^{\hat 2} &=& \left(1 - {\frac {a^2\cos^2\theta}{r^2}}\right)dr\wedge
 d\theta + \cdots,\\ F^{\hat 3} &=& -\,{\frac {a\sin\theta} {r^2}}\,cdt\wedge
 dr + {\frac {a\cos\theta} r}\,cdt\wedge d\theta + \sin\theta\,dr\wedge
 d\varphi \nonumber\\
&& +\,\left[r + {\frac{a^2}{2r}}\left(1+3\sin^2\theta\right)\right]
\cos\theta\,d\theta\wedge d\varphi + \cdots .
\end{eqnarray}
The dots denote terms of higher order in ${\frac 1r}$. Similarly, we find the
asymptotic connection
\begin{eqnarray}
\tilde{\Gamma}^{\hat{0}\hat{1}} &=& \left({\frac m {r^2}} + \cdots\right)cdt
+ \left(- {\frac {a\sin^2\theta} r} + \cdots\right)d\varphi, \label{G01KN}\\
\tilde{\Gamma}^{\hat{0}\hat{2}} &=& \left(-{\frac {a\sin\theta\cos\theta} r}
+ \cdots\right)d\varphi, \\
\tilde{\Gamma}^{\hat{0}\hat{3}} &=& \left(- {\frac {a\sin\theta}{r^2}} +
 \cdots \right)dr + \left({\frac {a\cos\theta} r} + \cdots\right)d\theta, \\
\tilde{\Gamma}^{\hat{1}\hat{2}} &=& \left(- {\frac {a^2\sin\theta\cos\theta}
{r^3}}
 + \cdots\right)dr + \left(- 1 + {\frac mr} + \cdots\right)d\theta, \\
\tilde{\Gamma}^{\hat{1}\hat{3}}&=& \left(- 1 + {\frac m r} + \cdots\right)
\sin\theta\,d\varphi,\\
\tilde{\Gamma}^{\hat{2}\hat{3}} &=& \left({\frac {2ma}{r^3}}
+ \cdots\right)\cos\theta\,cdt + \left(- 1 + 
\cdots\right)\cos\theta\,d\varphi.
\label{G23KN}
\end{eqnarray}
The resulting gravitational field momentum $\tilde{H}_\alpha$ is given by
\begin{eqnarray}
\kappa\tilde{H}_{\hat 0}&=& {\frac {2ma\cos\theta}{r^3}}\,cdt\wedge dr
- {\frac {a\sin\theta}{r}}\,cdt\wedge d\theta + \left(1 + {\frac m r}
\right)\cos\theta dr\wedge d\varphi\nonumber\\
&& + \left(-2r + 2m - {\frac {2a^2\sin^2\theta + q^2 - m^2} r}\right)
\sin\theta d\theta\wedge d\varphi + \cdots , \label{H0KN}\\
\kappa\tilde{H}_{\hat 1} &=& - {\frac {a\sin\theta} r}\,dr\wedge d\theta
- \left(1 - {\frac m r}\right)\cos\theta\,cdt\wedge d\varphi + \cdots,\\
\kappa\tilde{H}_{\hat 2} &=& {\frac {a\cos\theta} r}\,dr\wedge d\theta +
\left(1 - {\frac m r}\right)\sin\theta\,cdt\wedge d\varphi + \cdots,\\
\kappa\tilde{H}_{\hat 3} &=& - {\frac {a^2\sin\theta\cos\theta}{r^3}}
\,cdt\wedge dr - \left(1 - {\frac m r}\right)cdt\wedge d\theta
+ {\frac {a\sin\theta\cos\theta}r}\,dr\wedge d\varphi\nonumber\\
&& + a\sin^2\theta\left(-2 + {\frac m r}\right)d\theta\wedge d\varphi
+ \cdots. \label{H3KN}
\end{eqnarray}
The term $-2r\sin\theta d\theta\wedge d\varphi$ in (\ref{H0KN}) leads 
to a {\it divergent} total energy-momentum.

In order to find the regularizing connection (\ref{fc}), we can simply 
take limit of (\ref{G01KN})-(\ref{G23KN}) for $r\rightarrow\infty$. This
yields the result:
\begin{equation}\label{limGam}
\overline{{\Gamma}}{}^{\hat{0}\alpha}= 0,\qquad
\overline{{\Gamma}}{}^{\hat{1}\hat{2}}=-\,d\theta ,\qquad
\overline{{\Gamma}}{}^{\hat{1}\hat{3}}=-\sin\theta\,d\varphi,\qquad
\overline{{\Gamma}}^{\hat{2}\hat{3}} = -\cos\theta\,d\varphi.
\end{equation}
These components do not depend of time and radial coordinates, and we
extend by continuity the same values to all the spacetime manifold. We 
can verify that it is indeed a flat connection, i.e. its curvature vanishes
identically. We now take this flat connection and compute the
{\it relocalization} term $\frac{1}{2}\,\overline{{\Gamma}}
{}^{\mu\nu}\wedge\eta_{\alpha\mu\nu}$:
\begin{eqnarray}
\frac{1}{2}\,\overline{{\Gamma}}{}^{\mu\nu}\wedge\eta_{\hat{0}\mu\nu} &=&
-\,{\frac {a\sin\theta}r}\,cdt\wedge d\theta + \left(1 + {\frac m r}
\right)\cos\theta\,dr\wedge d\varphi\nonumber\\
&& - \,\left(2r + {\frac {a^2} r}\right)\sin\theta d\theta\wedge
d\varphi + \cdots ,\\
\frac{1}{2}\,\overline{{\Gamma}}{}^{\mu\nu}\wedge\eta_{\hat{1}\mu\nu} &=&
-\left(1 - {\frac m r}\right)\cos\theta\,cdt\wedge d\varphi + \cdots,\\
\frac{1}{2}\,\overline{{\Gamma}}{}^{\mu\nu}\wedge\eta_{\hat{2}\mu\nu} &=&
\left(1 - {\frac m r}\right)\sin\theta\,cdt\wedge d\varphi + \cdots,\\
\frac{1}{2}\,\overline{{\Gamma}}{}^{\mu\nu}\wedge\eta_{\hat{3}\mu\nu} &=&
- \left(1-{\frac m r}\right)cdt\wedge d\theta - a\left(1 -{\frac m r}\right)
\sin^2\theta\,d\theta\wedge d\varphi + \cdots.
\end{eqnarray}
Thus, the {\it relocalized momentum} (\ref{Hreg}) is found to be {\it 
regularized}:
\begin{eqnarray}
\widehat{H}_{\hat 0} &=& {\frac 1 \kappa}\Bigg[{\frac {2ma\cos\theta}
{r^3}}\,cdt\wedge dr 
+ \left(2m - {\frac {a^2\sin^2\theta + q^2 - m^2} r}\right)
\sin\theta d\theta\wedge d\varphi + \cdots\Bigg], \label{Hinf0} \\
\widehat{H}_{\hat 1} &=& {\frac 1 \kappa}\left[- {\frac {a\sin\theta} 
r}\,dr\wedge d\theta + \cdots\right],\\
\widehat{H}_{\hat 2} &=& {\frac 1 \kappa}\left[{\frac {a\cos\theta} 
r}\,dr\wedge  d\theta + \cdots\right],\\
\widehat{H}_{\hat 3} &=& {\frac 1 \kappa}\left[{\frac
{a\sin\theta\cos\theta}r}\,dr\wedge (d\varphi - {\frac a {r^2}}\,cdt) 
- a\sin^2\theta\,d\theta\wedge d\varphi + \cdots\right]. \label{Hinf3}
\end{eqnarray}
Finally, the corresponding conserved energy-momentum (\ref{Pa}) reads
\begin{equation}
\widehat{P}_\alpha=\left(Mc,\ 0,\ 0,\ -\frac{\pi ac^3}{8G}\right).
\end{equation}
Notice that at spatial infinity, the vector frame dual to
(\ref{KNcof0})-(\ref{KNcof3}) is of the form $e_{\hat
0}=\partial_t+O(r^{-2})$, $e_{\hat 1}=\partial_r+O(r^{-1})$, $e_{\hat
2}=\frac{1}{r}\partial_\theta+O(r^{-2})$ and $e_{\hat 3}=\frac{a}
{r}\sin\theta\partial_t+\frac{1}{r\sin\theta}\partial_\varphi+O(r^{-2})$. 
We see that $e_{\hat 3}$ has components along $\partial_t$ as well as along
$\partial_\varphi$. As a result, $P_{\hat 3}$ picks up a contribution that 
describes a ``momentum along the $\varphi$ direction", proportional to 
the rotation parameter $a$.

\subsubsection{A different choice of connection}

As we have already mentioned, the regularizing connection is not unique.
An alternative way to construct it is as follows. Take the Riemannian 
connection of the tetrad (\ref{KNcof0})-(\ref{KNcof3}) and put the mass 
and the charge parameters equal zero, $m = 0, q^2 = 0$. The result is
the flat connection with the components 
\begin{eqnarray}
\overline{{\Gamma}}{}^{\hat{0}\hat{1}}&=& -\frac {a'r \sin^2\theta}
{r^2+a'^2\cos^2\theta}\,d\varphi,
\qquad \overline{{\Gamma}}{}^{\hat{0}\hat{2}}=-{\frac {a'\,\sin \theta\cos 
\theta\sqrt {{r}^2+a'^2}  }{r^2+a'^2 \cos^2\theta}}\,d\varphi ,\\
\overline{{\Gamma}}{}^{\hat{0}\hat{3}}&=&-{\frac {a' r\sin \theta}
{ \left( {r}^2+a'^2 \cos^2\theta
 \right) \sqrt {{r}^2+a'^2}}}\,dr+\frac{a'\cos\theta \sqrt{r^2+a'^2}}
{r^2+a'^2\cos^2\theta}\,d\theta,\\
\overline{{\Gamma}}{}^{\hat{1}\hat{2}}&=&-\frac{a'^2\sin\theta\cos\theta}
{(r^2+a'^2 \cos^2\theta) \sqrt {{r}^2+a'^2}}\,dr -\frac{r\sqrt{r^2+a'^2}}
{r^2+a'^2\cos^2\theta}\,d\theta , \\
\overline{{\Gamma}}{}^{\hat{1}\hat{3}}&=&-\frac{r\sin\theta\sqrt{r^2+a'^2}}
{r^2+a'^2\cos^2\theta}\,d\varphi,\qquad
\overline{{\Gamma}}^{\hat{2}\hat{3}} = -\frac{\left(r^2+a'^2\right)\cos\theta}
{r^2+a'^2\cos^2\theta}\,d\varphi.
\end{eqnarray}
We have put the rotation parameter different from the original Kerr-Newman
one, $a'\neq a$, in order to stress the different nature of this connection. 
In the limit of $r\rightarrow\infty$ this connection tends to (\ref{limGam}),
so the condition $\Delta\Gamma^{\mu\nu}\vline_{\partial S}= 0$ is satisfied. 
For the new choice, the relocalization yields the regularized momentum  
(\ref{Hreg}) with the asymptotic $r\rightarrow\infty$ behavior
\begin{eqnarray}
\widehat{H}_{\hat 0} &=& \frac{2m}{\kappa}\sin\theta\, d\theta\wedge 
d\varphi ,\\
\widehat{H}_{\hat 1} &=&\widehat{H}_{\hat 2} = 0,\\
\widehat{H}_{\hat 3} &=& (a'-a)\sin^2\theta\,d\theta\wedge d\varphi.
\end{eqnarray}
Accordingly, the regularized total energy-momentum (\ref{Pa}) reads
\begin{equation}
\widehat{P}_\alpha=\left(Mc,\ 0,\ 0,\ \frac{\pi c^3}{8G}(a'-a)\right).
\end{equation}
This example demonstrates that the total conserved quantities obtained 
with the help of the regularization from the initially divergent momentum 
are sensitive to the choice of flat connection $\overline{\Gamma}$.

\subsubsection{Rotated regular coframe}

The connection (\ref{limGam}), (\ref{flatGam}) is flat. As a result, 
we can write it as $\overline{\Gamma}_\alpha{}^\beta = (\Lambda^{-1})
{}^\beta{}_\gamma d\Lambda^\gamma{}_\alpha$ with some Lorentz matrix. 
With the help of the latter, we can rotate the original coframe 
(\ref{KNcof0})-(\ref{KNcof3}) 
to a new tetrad $\vartheta'{}^\alpha = \Lambda^\alpha{}_\beta\,
\vartheta^\beta$. 
Explicitly, this local Lorentz transformation can be chosen as:
\begin{eqnarray}
\vartheta'{}^{\hat 0} &=& \vartheta^{\hat 0}, \label{theta'0}\\
\vartheta'{}^{\hat 1} &=& \vartheta^{\hat 1}\cos\varphi\sin\theta +
\vartheta^{\hat 2}\cos\varphi\cos\theta
- \vartheta^{\hat 3}\sin\varphi,\\
\vartheta'{}^{\hat 2} &=& - \vartheta^{\hat 1}\cos\theta +
\vartheta^{\hat 2}\sin\theta,\\
\vartheta'{}^{\hat 3} &=& \vartheta^{\hat 1}\sin\varphi\sin\theta +
\vartheta^{\hat 2}\sin\varphi\cos\theta
+ \vartheta^{\hat 3}\cos\varphi. \label{theta'3}
\end{eqnarray}
We can compute the conserved quantities in the new frame. First of all, 
we find that the Riemannian connection for this coframe has vanishing limit 
at spatial infinity. This means that no regularization is needed in this 
case, as expected. Furthermore, the limit value of the field momentum 2-form 
$\widehat{H}'_\alpha = \tilde{H}'_\alpha$ at spatial infinity is given by
\begin{equation}
\widehat{\overline{H'}}_{\alpha}
=\left(\frac{2m}{\kappa}\sin\theta\,d\theta\wedge d\varphi,\,\frac{a}
{\kappa}\sin^2\theta\sin\varphi\,d\theta\wedge d\varphi,\,0,\,-\frac{a}
{\kappa}\sin^2\theta\cos\varphi\,d\theta\wedge d\varphi\right) .\label{H'inf}
\end{equation}
The corresponding conserved total energy-momentum 
$\widehat{P}'_\alpha=(Mc,0,0,0)$ and hence we can interpret  $M$ as 
the total mass of the system (since $e_{\hat 0} =\partial_t +O(1/r)$ at 
spatial infinity). Additionally, we can verify that the regularized 
momentum at spatial infinity is indeed covariant under local Lorentz
transformations at infinity. From (\ref{Hinf0})-(\ref{Hinf3}) we find, 
at a spatial infinity,
\begin{equation}
\widehat{\overline{H}}_{\alpha} =\left(2m\sin\theta,\,0,\,0,\,-a\sin^2\theta
\right)\,\frac{1} {\kappa}d\theta\wedge d\varphi .\label{Hinf}
\end{equation}
It is then straightforward to confirm that (\ref{H'inf}) and (\ref{Hinf}) are
related by $\widehat{\overline{H'}}_{\alpha}=(\Lambda^{-1})^\beta_{\
\alpha}\widehat{\overline{H}}_{\alpha}$.

This example demonstrates that it is the asymptotics  \textit{of  the 
Riemannian connection 1-form at spatial infinity} ($\overline{\Gamma}_\alpha{}
^\beta$), and \textit{not of the anholonomity 2-form} $\overline{F}^\alpha$,
that is crucial for determining the need of the regularization procedure. 
Indeed, for the new frame (\ref{theta'0})-(\ref{theta'3}) we find the
nontrivial limit of anholonomity $\overline{F}^\alpha=(-2\sin\theta\cos
\theta\,d\theta\wedge d\varphi,0,0,0)$ while, by construction, the limit
of the Riemannian connection is trivial $\overline{\Gamma}_\alpha{}^\beta
= 0$. Since the Riemannian curvature is asymptotically zero, the vanishing
connection means that the corresponding frame is asymptotically inertial, 
whereas a nonvanishing Lorentz connection can be interpreted in terms of the 
acceleration and rotation of observers associated to that frame. Thus, we 
can say that the regularization procedure ``corrects" the conserved 
quantities by removing ``noninertial terms" at spatial infinity. If a frame
is already asymptotically inertial (that is, such that its Riemannian 
connection vanishes asymptotically) then no regularization is necessary.

\subsection{Example 2: configuration with mass, charge and scalar field}

As another example, let us consider the exact spherically symmetric solution
for the coupled system of the scalar, electromagnetic and gravitational field.
We take the coframe as \cite{scalar1,scalar2}
\begin{equation}
\vartheta^{\hat 0}=\sqrt{f}\,cd\tau,\
\vartheta^{\hat 1}=\frac{1}{\sqrt{f}}\,d\rho,\ 
\vartheta^{\hat 2}=\sqrt{h}\,d\theta,\
\vartheta^{\hat 3}=\sqrt{h}\sin\theta\,d\varphi.\label{Svta}
\end{equation} 
Here the time and the radial coordinates $(\tau,\rho)$ are (for vanishing 
scalar field) different from the above $(t,r)$, and the functions depend 
on the variable $y=2m/\rho$ only:  
\begin{eqnarray}
f&:=&\frac{(1-y^2)^\mu}{\left[k^2(1+y)^\mu-(1-y)^\mu\right]^2}, \\
h&:=&\rho^2(1-y^2)^{1-\mu}\left[k^2(1+y)^\mu-(1-y)^\mu\right]^2.
\end{eqnarray} 
The massless scalar field $\Phi$ and the electromagnetic field
strength 2-form $F$ are, respectively:
\begin{eqnarray}
\Phi &=& \sqrt{{1-\mu^2\over 2}}\,
\log\left\vert{1 - y\over 1 + y}\right\vert,\label{phi2}\\
F &=& \pm {\frac {4m\mu} h}\sqrt{\frac {k^2c^2} {4\pi\varepsilon_0 G}}
\,\vartheta^{\hat 1}\wedge\vartheta^{\hat 0}.\label{max2}
\end{eqnarray}
Here $\varepsilon_0$ is the electric constant of the vacuum.
The solution (\ref{Svta}), (\ref{phi2}), (\ref{max2})
depends on three arbitrary integration constants: $\mu$ and $k^2$ 
are dimensionless and $m$ has the dimension of length. The total 
electric charge $Q$ of the system is obtained as the integral of
the electric excitation 2-form $\cal D$ over the boundary 2-sphere of the 
spatial volume, $Q = \int_{\partial S} {\cal D}$. Explicitly, we find 
\begin{equation}
{\cal D} = H = \varepsilon_0c\,{}^\star\!F = \pm 4m\mu\varepsilon_0c
\sqrt{\frac {k^2c^2} {4\pi\varepsilon_0 G}}
\,\sin\theta d\theta\wedge d\varphi. 
\end{equation}
Accordingly, the total charge is 
\begin{equation}
Q = \pm 4m\mu\sqrt{4\pi\varepsilon_0c^4\,k^2/G}.
\end{equation}
As we see, all the three integration constants contribute to the
{\it total electric charge} of the solution. 

Let us now calculate the {\it total mass} of the solution. We begin with
the analysis of the asymptotic behavior of the coframe. For $k^2> 1$,
we find  
\begin{eqnarray}
\vartheta^{\hat 0}&=& \left[\frac{1}{k^2-1}-\frac{2\mu m(1+k^2)}
{(k^2-1)^2\rho}+\cdots\right]\,d\tau, \\
\vartheta^{\hat 1}&=&\left[(k^2-1)+\frac{2\mu m(1+k^2)}
{\rho}+\cdots\right]\,d\rho,\\
\vartheta^{\hat 2}&=&\left[(k^2-1)\rho+2\mu m(1+k^2)+\frac{2m^2(k^2-1)
(\mu^2-1)}{\rho}+\cdots\right]\,d\theta, \\
\vartheta^{\hat 3}&=&\left[(k^2-1)\rho+2\mu m(1+k^2)+\frac{2m^2(k^2-1)
(\mu^2-1)}{\rho}+\cdots\right]\sin\theta\,d\varphi .
\end{eqnarray} 
The asymptotic non-regularized gravitational field momentum then reads
\begin{eqnarray}
\kappa\tilde{H}_{\hat 0}&=&\left[(k^2-1)+\frac{2\mu m(1+k^2)}
{\rho}\right]\cos\theta\,  d\rho\wedge d\varphi \nonumber\\
&&-\left[2(k^2-1)\rho-\frac{4m^2\mu^2(k^2-1)}{\rho}\right]\sin\theta\,  
d\theta\wedge d\varphi + \cdots , \label{H0KNY}\\
\kappa\tilde{H}_{\hat 1} &=& -\left[\frac{1}{k^2-1}-\frac{2\mu m(1+k^2)}
{(k^2-1)^2\rho}\right]\cos\theta\,d\tau\wedge d\varphi + \cdots,\\
\kappa\tilde{H}_{\hat 2} &=&\left[\frac{1}{k^2-1}-\frac{2\mu m(1+k^2)}
{(k^2-1)^2\rho}\right]\sin\theta\,d\tau\wedge d\varphi + \cdots,\\
\kappa\tilde{H}_{\hat 3} &=&-\left[\frac{1}{k^2-1}-\frac{2\mu m(1+k^2)}
{(k^2-1)^2\rho}\right]d\tau\wedge d\theta+ \cdots. \label{H3KN2}
\end{eqnarray}
In view of the divergent term in (\ref{H0KNY}), regularization is 
needed. For the limiting connection $\overline{\Gamma}$, we again obtain 
(\ref{limGam}). So, we can perform the regularization with the same flat
connection used before. The final regularized gravitational field momentum 
then has 
the asymptotics
\begin{eqnarray}
\kappa\widehat{H}_{\hat 0} &=& \left[4\mu m(1+k^2)+\frac{4m^2(2\mu^2-1)
(k^2-1)}{\rho}+\cdots\right]\sin\theta\,d\theta\wedge d\varphi , \\
\kappa\widehat{H}_{\hat 1} &=& 0,\\
\kappa\widehat{H}_{\hat 2} &=& \frac{2m^2}{(k^2-1)\rho^2}\sin\theta
\,d\tau\wedge d\varphi+\cdots,\\
\kappa\widehat{H}_{\hat 3} &=& -\frac{2m^2}{(k^2-1)\rho^2}
\,d\tau\wedge d\theta + \cdots.
\end{eqnarray} 

This yields, for the regularized total energy-momentum, 
\begin{equation}
P_\alpha=\left(2\mu mc^3(1+k^2)/G,\,0,\,0,\,0\right).
\end{equation}
Consequently, the total gravitating mass is $M=2\mu mc^2(1+k^2)/G$. To put 
it differently, we find the combination of the parameters in terms of the 
total mass: $2\mu m(1+k^2) = GM/c^2$. For the case $k^2<1$ all the above 
quantities change sign except for $\overline{\Gamma}$ that remains the same, 
so we obtain $M=-2\mu mc^2(1+k^2)/G$. The requirement of positive total 
energy then implies that we have to choose $\mu m>0$ for $k^2>1$ and 
$\mu m<0$ for $k^2<1$.

\section{Comparing tetrad and metric formulations}

There exists a natural relation between the tetrad formulation and the 
traditional metric formulation of GR. It is based on the fact that the
usual theory is actually a tetrad theory but just considered with respect
to the holonomic coordinate frame $dx^i$. Accordingly, the basic quantities
in the tetrad and in the metric formulations are related by means of the
linear transformation $dx^i = L^i{}_\alpha\vartheta^\alpha$ from the 
anholonomic frame $\vartheta^\alpha$ to the holonomic frame $dx^i$. The 
components of the corresponding matrix are evidently just the coefficients
of the coframe, i.e., $L^i{}_\alpha = h^i_\alpha$.

It is straightforward to generalize the transformation formulas 
(\ref{Vprime})-(\ref{Hprime}) by extending them from the Lorentz matrices
$\Lambda$ to the linear matrices $L$. For the field momentum we then find
the transformation
\begin{equation}
\tilde{H}'_i(dx) = (L^{-1})^\alpha{}_i\tilde{H}_\alpha(\vartheta) -{\frac 
1{2\kappa}}(L^{-1})^\alpha{}_i (L^{-1})^\nu{}_jdL^j{}_\mu\wedge\eta_\alpha
{}^\mu{}_\nu\,.\label{HHh}
\end{equation}

Now let us compare the explicit components of the field momentum 
(\ref{Htilde}) in the {\it anholonomic} frame with its components in the 
{\it holonomic} frame. First we recall that $\eta_{\alpha\beta\gamma} = 
\sqrt{-g}h^i_\alpha h^j_\beta h^k_\gamma\,\epsilon_{ijk}$ and $\eta_{
\alpha\beta} = \sqrt{-g}h^i_\alpha h^j_\beta\,\epsilon_{ij}$, where 
$\epsilon_{ijk} = \epsilon_{ijkl}\,dx^l$ and $\epsilon_{ij} = {\frac 12}
\epsilon_{ijkl}\,dx^k\wedge dx^l$ with the numeric Levi-Civita symbol 
$\epsilon_{ijkl}$. Then directly from (\ref{Htilde}) we find
\begin{equation}\label{MollU}
\tilde{H}_\alpha = h^k_\alpha\,{\cal U}_k{}^{ij}\,\epsilon_{ij},\qquad 
{\cal U}_k{}^{ij} = {\frac {\sqrt{-g}}{2\kappa}}\left(\gamma_k{}^{ij} +
\delta^i_k\gamma_l{}^{jl} - \delta^j_k\gamma_l{}^{il}\right),
\end{equation}
where $\gamma_{ij}{}^k := h^\alpha_j\left(\partial_ih^k_\alpha +
\tilde{\Gamma}^k_{il}h^l_\alpha\right)$ is defined with the help of the
usual Christoffel connection $\tilde{\Gamma}^k_{il} = {\frac 12}g^{kj}
\left(\partial_ig_{lj} + \partial_lg_{ij} - \partial_jg_{kl}\right)$. 
We thus see that the field momentum $\tilde{H}_\alpha$ in the anholonomic 
frame is expressed in terms of the M\o{}ller superpotential ${\cal 
U}_k{}^{ij}$ \cite{Moller}.

Now, directly from (\ref{HHh}), we find the same field momentum in the 
holonomic frame:
\begin{eqnarray}
\tilde{H}'_k(dx) &=& h^\alpha_k\tilde{H}_\alpha(\vartheta) -{\frac 
1{2\kappa}}h^\alpha_k h^\nu_jdh^j_\mu\wedge\eta_\alpha{}^\mu{}_\nu\\
&=& V_k{}^{ij}\,\epsilon_{ij},
\end{eqnarray}
with 
\begin{equation}
 V_k{}^{ij} = {\frac {g_{kl}}{4\kappa
\sqrt{-g}}}\,\partial_m\!\left[(-g)\left(g^{il}g^{jm} - g^{im}g^{jl}\right)
\right].\label{EinsU}
\end{equation} 
We thus recover the components of the Freud superpotential \cite{Freud} for 
the Einstein \cite{Ein} energy-momentum. 

This result shows that the well known energy-momentum complexes of M\o{}ller
and of Einstein(-Freud) are actually different faces (anholonomic and 
holonomic, respectively) of {\it one and the same object}. This fact seems to
be not noticed in the previous literature.

\section{Final remarks}

In this paper we have studied, within the tetrad formulation of GR, the 
covariance properties of various quantities describing the local and global 
energy-momentum content of gravitating systems. Our main result is given given 
by the formulas (\ref{Vprime})-(\ref{Hprime}) that describe the transformation 
laws of the Lagrangian, the energy-momentum current and the field momentum
(the respective ``superpotential") under local Lorentz transformations 
of a frame. The total energy-momentum does not ``feel" the changes of frame
inside the compact spatial region bounded by $\partial S$, but it is
sensitive to the local Lorentz transformation of a frame at spatial infinity. 

In general, the total conserved energy-momentum $\widetilde{P}_\alpha$ 
corresponding to $\widetilde{H}_\alpha$ does not transform covariantly under 
a change of frame. However, for local Lorentz transformations which become 
global at spatial infinity, the total energy-momentum transforms covariantly
as a Lorentz vector.

When the Riemannian connection of a given frame is nontrivial at spatial 
infinity, the total energy-momentum can diverge and a regularization is 
needed. We have shown that it is possible to regularize the energy-momentum
with the help of a relocalization defined by the flat connection $\overline{
\Gamma}$. In addition, the regularized field momentum and canonical 
energy-momentum current turn out to be covariant under local Lorentz 
transformations. Note that relocalization of the energy-momentum currents 
was also recently discussed in \cite{Hannibal,Itin2}.

It is thus possible to define global conserved quantities in any orthonormal 
frame, independently of any extra symmetry of spacetime (Killing vectors), 
cf. \cite{CN99} and references therein. Alternatively, for a given vector 
field $\xi$ we can define a global conserved quantity by
\begin{equation}
 Q:=\int_{\partial S}(\xi\rfloor\vartheta^\alpha)\wedge 
H_\alpha=\int_{\partial S}\xi^\alpha H_\alpha.
\end{equation} 
The global conserved quantity $Q$ is invariant (for any given vector field
$\xi$) under both the general coordinate transformations and under local 
Lorentz transformations of a frame when we use the regularized field 
momentum $\widehat{H}_\alpha$ above. 

\bigskip
{\bf Acknowledgements}. The authors would like to thank J.G. Pereira for 
useful discussions and for his hospitality at IFT-UNESP. This work was 
supported by FAPESP (for YNO) and by CNPq (for GFR). We also thank 
F.W. Hehl for the comments on our paper.

\end{document}